\documentclass[floats,aps,prb,showpacs,twocolumn,superscriptaddress]{revtex4}
\usepackage{graphicx}

\usepackage{amsmath,amssymb,latexsym,amsfonts,subfigure,bbm}

\begin{document}
\title{Statistics of Transmission Eigenvalues in Two-Dimensional Quantum Cavities:\\Ballistic
  versus Stochastic Scattering}

\author{Stefan Rotter}
\affiliation{Institute for Theoretical Physics,
             Vienna University of Technology, A-1040 Vienna, Austria, EU}
\affiliation{Department of Applied Physics, Yale University, New Haven, CT 06520, USA}
\author{Florian Aigner}
\affiliation{Institute for Theoretical Physics,
             Vienna University of Technology, A-1040 Vienna, Austria, EU}
\author{Joachim Burgd\"orfer}
\affiliation{Institute for Theoretical Physics,
             Vienna University of Technology, A-1040 Vienna, Austria, EU}

\date{\today} 

\begin{abstract}
We investigate the statistical distribution of transmission eigenvalues
in phase-coherent transport through quantum dots. In two-dimensional 
ab-initio simulations for both clean and disordered two-dimensional cavities, 
we find markedly different quantum-to-classical crossover scenarios for these
two cases. In particular, we observe the emergence of ``noiseless scattering
states'' in clean cavities, irrespective of sharp-edged entrance and exit lead mouths. 
We find the onset of these ''classical'' states to be largely independent of the
cavity's classical chaoticity, but very sensitive with respect to bulk disorder.
Our results suggest that for weakly disordered cavities the transmission
eigenvalue distribution
is determined both by scattering at the disorder potential and the cavity walls. To
properly account for this intermediate parameter regime we introduce a hybrid
crossover scheme which combines previous models that are valid 
in the ballisic and the stochastic limit, respectively.
\end{abstract}

\maketitle

Shot noise, i.e.~the fluctuations of the current due to the statistical
nature of charge transport, has recently become an intensively studied subject-matter in
the field of mesoscopic physics. Being first investigated on a macroscale now almost
a century ago,\cite{schottky} the interest in this phenomenon has
recently witnessed a revival (see Ref.~\onlinecite{been1} 
for an introduction to this topic and 
Ref.~\onlinecite{blanter1} for an extensive review). 
On the experimental side, modern
semiconductor fabrication techniques have allowed
for high-precision experiments of quantum shot noise.
\cite{steinbach,henny,schoelkopf,oberholzer0,oberholzer1,reulet}
On the theoretical side it was demonstrated that these measurements
allow to extract detailed information on microscopic transport mechanisms 
which are difficult to access otherwise.
\cite{khlus,doro,imry,agam,blanter2,blanter3,tworz,jacquod,silvestrov,aigner,sukhorukov,dejong,sim,naz,macu,bulashenko2,whitney,falko,rahav,nagaev}\\\cite{hekking,marconcini,gopar}

Since shot noise on the mesoscopic scale is due to the quantum
(probabilistic) nature of transport, a suppression of shot noise has been 
predicted \cite{agam} as transport becomes more classical (or deterministic), 
i.e.~when the
ratio of the Fermi wavelength $\lambda_F$ to the linear cavity size $L$
vanishes, $\lambda_F/L\rightarrow 0$.
Whereas this prediction has meanwhile been numerically
\cite{tworz,jacquod,naz,sim,macu,aigner,whitney,rahav} as well as
experimentally \cite{oberholzer1} confirmed, it is still a subject of debate
how to identify signatures of the different 
sources of noise in this quantum-to-classical crossover. 
On the theoretical side, different analytical predictions describe 
the quantum-to-classical crossover for
cavities with ballistic scattering (off smooth potential or boundary profiles) 
\cite{agam,silvestrov,tworz,jacquod}
or with disorder scattering (off short-range impurities or rough
boundaries).\cite{sukhorukov,blanter2}
In ballistic dots the crossover to the noiseless classical regime 
is anticipated to be mediated by ``noiseless
scattering states''.\cite{silvestrov} The 
separation of phase space in
noiseless classical (i.e.~deterministic) and noisy quantum channels
is in sharp contrast to the case of cavities with bulk disorder
where all transporting channels are expected to contribute
to shot noise.\cite{falko,jacquod,sukhorukov,whitney,aigner,feist} Testing the
validity of these theories has turned out to be a major challenge:
Measurements suffer from limited accuracy and seem to be able to explore 
only the onset of the quantum-to-classical crossover where different
models are difficult to distinguish from each other.\cite{oberholzer1}
Also numerical simulations for two-dimensional (2D) transport 
\cite{sim,naz,macu,marconcini,aigner} suffer from slow convergence for
 $\lambda_F/L\rightarrow 0$, which reason has prevented a detailed test of 
differing predictions in that limit. To circumvent this problem an
open dynamical kicked rotator model has recently been
used to mimic chaotic as well as stochastic scattering in a 1D system.
\cite{tworz,jacquod,whitney,rahav}
While being computationally more easily tractable, especially in the 
semiclassical regime of small $\lambda_F$, these stroboscopic
models do, however, not
fully incorporate features of 2D transport which contribute
significantly to the shot noise -- as, e.g., whispering gallery
modes\cite{naz} and an accurate description of diffraction 
at the dot openings or at a bulk disorder
potential.\cite{aigner,marconcini}

\begin{figure}[!b]
      \includegraphics[angle=0,width=85mm]{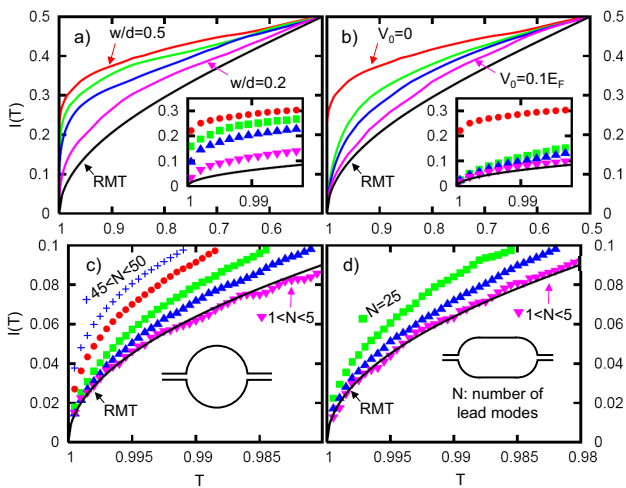}
      \caption{(Color online) Integrated distribution function
        of transmission eigenvalues, $I(T)$.
        Top row: Rectangular billard with tunable shutters (see inset
        Fig.~\ref{fig:2}). 
        (a) Crossover from large to small shutter openings (at zero disorder, $V_0=0$): I(T) for 
        $w/d = 0.5, 0.4, 0.3$, and $0.2$ (top to bottom).
        (b) Crossover from clean to disordered samples (at half-opening, $w/d=0.5$): I(T) for different
        disorder potentials $V_0 = 0, 0.03 E_F, 0.05 E_F$, and $0.1 E_F$ (top
        to bottom). The pronounced difference between (a) and (b)
        near $T=1$ is highlighted in the
        insets. Bottom row: $I(T)$ for the circular
        (c) and the stadium-shaped geometry (d) for different values of $k_F$.
        In (c) we average over the intervals  $45\!<\!N\!<\!50$, 
        $30\!<\!N\!<\!35$, $20\!<\!N\!<\!25$, $10\!<\!N\!<\!15$ 
        and $1\!<\!N\!<\!5$ (top to bottom), 
        in (d) $N=25$, $10\!<\!N\!<\!15$, $1\!<\!N\!<5$ (top to bottom),
        where $N$ is the number of open lead modes.
        The RMT limit is indicated by black lines.}
        \label{fig:1}
\end{figure}

The aim of the present communication is to provide such a
2D transport simulation in the quantum-to-classical
crossover regime. Our calculations are performed
within the framework of the modular recursive
Green's function method (MRGM).\cite{rotter1} In our
single-particle model, effects of finite temperature and electron-electron 
interaction are neglected. 
Finite temperatures would lead to a cross-over from shot noise to
thermal noise and inelastic electron-electron interactions would increase the 
noise.\cite{dejong,blanter2} The effect of both mechanisms can however be
controlled in the experiment by reducing temperature and system size down to
a regime where inelastic scattering sources can be neglected.\cite{steinbach,henny,oberholzer0}
We study cavities with $N$ open channels in each of the two attached
leads of equal width (injection from the left) and characterize
the transport problem by the transmission ($t$) and 
reflection ($r$) matrices of dimension $N\times N$.
Following the Landauer-B\"uttiker theory, the transmission eigenvalues 
$T_n$ of the matrix $t^\dagger t$ 
determine the average current, 
$\langle I\rangle=\Delta\mu\sum_n T_n$ and the shot noise power,\cite{khlus}
$S\equiv\langle\langle I^2 \rangle\rangle=\Delta\mu\sum_n T_n(1-T_n)$ 
(assuming $e=h=1$,
and a chemical potential difference $\Delta\mu$ between the two leads).  
Since also all higher cumulants of the current, 
$\langle\langle I^m\rangle\rangle$, are determined by the eigenvalues $T_n$,
knowledge of the distribution function of the eigenvalues, $P(T)$, allows to
obtain the full counting statistics of the transport problem.\cite{lee} 
The distribution $P(T)$ will be at the center of our attention 
in the present article as different mechanisms of transport leave conspicuous 
signatures on its functional form. It was first pointed out for the case of a diffusive wire, that
this system's eigenvalue distribution function features a bimodal distribution with
maxima near values of very high ($T\approx 1$) and very low ($T\approx 0$)
transmission, respectively.\cite{doro,imry} The effects of this feature on the suppression
of noise have been predicted \cite{nagaev} and were successfully
measured in the experiment,\cite{steinbach,schoelkopf,henny} as well as
simulated numerically.\cite{macu}
For classically chaotic rather than diffusive systems with $N\gg 1$ and time reversal symmetry, 
random matrix theory (RMT) predicts $P(T)$ to follow also a bimodal 
universal form,\cite{jalabert}
\begin{equation}
P_{RMT}(T) = \pi^{-1}\,[T(1-T)]^{-1/2} %\frac{1}{\pi \sqrt{T (1-T)}}
,\quad  T \in [0,1]\,.
\label{prmt}
\end{equation}
For ballistic cavities the quantum-to-classical crossover of this eigenvalue
distribution is predicted to proceed via ``noiseless states'',
\cite{silvestrov,jacquod}
\begin{equation}
P^\alpha(T) = \alpha\, P_{RMT}(T) \!+\! (1\!-\!\alpha)\,\left[
  \delta(T) + \delta(1\!-\!T) \right]/2\,.
\label{cross1}
\end{equation}
Noiseless (or deterministic)
 transport channels with eigenvalues $0$ or $1$ and weight $(1-\alpha)$
[represented by the 
last two terms in Eq.~(\ref{cross1})] 
are expected to appear as soon as classical transmission bands \cite{wirtz1}
in phase space can be resolved by the quantum scattering 
process.\cite{silvestrov} For a chaotic system 
% with an exponential area distribution of transmission bands, 
the continuous crossover parameter $\alpha\in(0,1)$ was predicted\cite{agam} to scale as  
$\alpha=\exp(-\tau_E/\tau_D)$ with $\tau_D$ being the 
dwell time and $\tau_E$ the Ehrenfest time in the cavity. 
The latter estimates the time that it takes for a well-localized quantum 
wave packet to spread to the size $d$ of the cavity
($d\approx \sqrt{A}$ with $A$ the area of the dot) due to diverging 
classical trajectories. 
With the help of the Lyapunov exponent $\Lambda$, which measures
the rate of 
this divergence, the Ehrenfest time is typically estimated as:\cite{zas}
$\tau_E\approx\Lambda^{-1}\ln(d/\lambda_F)$.
Note, however, that this estimate requires corrections for regular or weakly chaotic
systems.\cite{aigner}  

In the presence of a uniform disorder with a correlation length smaller than the
electron wavelength $\lambda_F$ (``short-range bulk disorder'') the formation of 
noiseless states is suppressed by stochastic scattering. Also the trajectory-based
concept of the Ehrenfest-time as a crossover parameter breaks down here, leading to 
a different crossover form,\cite{sukhorukov}
\begin{equation}
P^\beta (T) = P_{RMT}(T)\,\ln\beta
\int_{-1}^1 du \,
\frac{(1-u)^2 |u|^{-(1+2\ln \beta)}}{4 T u-(1+u)^2}\,.
\label{cross2}
\end{equation}
The crossover parameter $\beta=\exp(-\tau_Q/\tau_D) \in(0,1)$ features the  
characteristic scattering time $\tau_Q$ which measures the
time within which an initially well-localized 
wave packet is stochastically scattered into random
direction. Note that the {\it stochastic} crossover, Eq.~(\ref{cross2}), 
interpolates between the same limiting cases $P_{RMT}$ 
(for $\beta \rightarrow 1$) 
and $P_{cl}=[\delta(T)+\delta(1-T)]/2$ (for $\beta \rightarrow 0$, 
i.e.~vanishing disorder) 
as the {\it ballistic} crossover in Eq.~(\ref{cross1}).

\begin{figure}[!b]
      \includegraphics[angle=0,width=85mm]{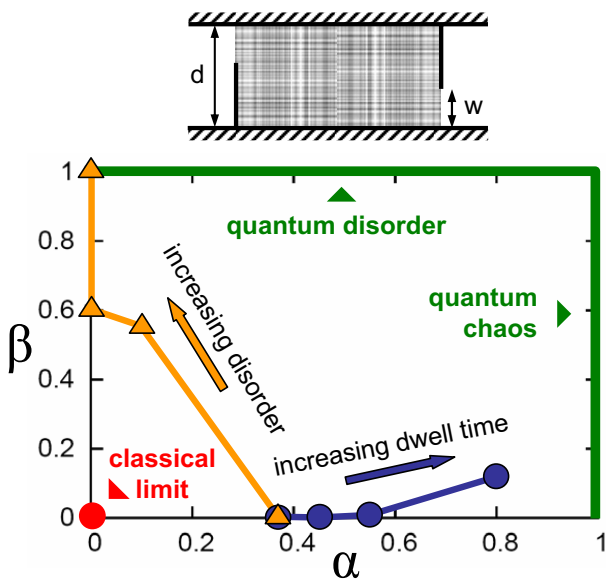}
      \caption{(Color online) Evolution of the crossover trajectories
        in the parameter space $(\alpha,\beta)$, obtained by fitting 
        Eq.~(\ref{cross3}) to the data displayed in Fig.~\ref{fig:1}a,b. 
%        Clasical limit: $P_{cl}$, quantum limit: $P_{RMT}$. 
        Starting parameters of both trajectories: $w/d=0.5, V_0=0$.
        Trajectory 1 for fixed $w/d=0.5$ (orange triangles): $V_0=0.03E_F$, $0.05E_F$ and $0.1E_F$. 
        Trajectory 2 for fixed $V_0=0$ (blue circles): $w/d=0.4$, $0.3$ and $0.2$.
        Inset: Rectangular cavity with tunable shutter openings and disorder 
        strength.}
        \label{fig:2}
\end{figure}

We now search for signatures of these two crossover scenarios 
in the numerical results for $P(T)$. To this end we calculate transport
through a rectangular cavity (see inset Fig.~\ref{fig:2}) 
with area $d\times2d$ and two tunable
openings (''shutters'') of width $w$ (inspired by recent shot noise
experiments\cite{oberholzer1}). 
The cavity interior contains 
a static bulk disorder potential $V$ with a
mean value $\langle V\rangle=0$ and a correlation function $\langle
V(x)V(x+a)\rangle=\langle V^2\rangle\exp(-a/l_C)$. The correlation length
$l_C$ is smaller than the Fermi wavelength,
$l_C/\lambda_F\approx 0.12$, and the potential strength $V_0=\sqrt{\langle
  V^2\rangle}$ corresponds to moderate disorder, $V_0\in[0,0.1]\times E_F$
(for details on the disorder potential see Ref.~\onlinecite{aigner}). 
In the limit of vanishing disorder strength
($V_0\rightarrow0$) the motion inside the rectangular 
cavity becomes completely regular.
%We calculate the scattering matrix at $400$ aequidistant points in the energy 
%range of $k_F \in [40.1, 40.85] \times \pi /d$.

We calculate 400 equidistant points in the interval
$k_F\in [40.1, 40.85] \times \pi /d$. In order to better resolve the
behavior of $P(T)$ near $T=1$ we plot the integrated eigenvalue distribution 
\cite{jacquod,sukhorukov} $I(T)=\int_{T}^1 P(\tau)d\tau$. 
For cavity parameters favorable to 
the appearance of noiseless scattering channels, i.e.~vanishing 
disorder ($V_0=0$), large openings ($w=d/2$), and large $k_F$,
we find that $I(T)$ features a very pronounced offset at
$T\approx 1$ (see Fig.~\ref{fig:1}a), corresponding to a statistically 
significant portion of effectively noiseless eigenvalues $T>0.999$. 
To verify whether these 
``classical'' transmission eigenvalues are indeed due to {\it direct 
scattering processes}, we control their
weight by gradually decreasing the cavity openings $w$ (Fig.~\ref{fig:1}a). 
Reducing $w$ decreases the offset and gradually 
shifts the distribution $P(T)$ towards
its RMT-limt [Eq.~(\ref{prmt})] for $w\rightarrow 0$. 
This behavior is all the more interesting as 
our sharp cavity openings do give rise to diffractive scattering \cite{wirtz1,aigner} 
which might suppress the formation of noiseless states. 
Our observation suggests, however, that noiseless transmission can still occur 
when scattering states effectively bypass any diffractive corners.
\cite{sukhorukov,whitney} To further test this hypothesis we now gradually
turn on the 
bulk disorder strength up to values of $V_0=0.1\!\times\! E_F$. Bulk disorder 
cannot be bypassed by any transmitting state and should therefore 
destroy the noiseless channels and, consequently, 
the offset in $I(T)$. We find that already a small disorder potential 
($V_0= 0.03\times E_F$) suppresses the offset in $I(T)$ 
entirely (Fig.~\ref{fig:1}b). With higher values of $V_0$ we reach the RMT-limit for $I(T)$. 
The striking difference 
between the ballistic (Fig.~\ref{fig:1}a) and the stochastic crossover (Fig.~\ref{fig:1}b) is best
visualized by zooming into the distribution $I(T)$ at values close to $T=1$
where the gradual vs.~''sudden'' suppression of the offset becomes most apparent 
(see insets in Figs.~\ref{fig:1}a,b). 
%For ballistic 
%cavities noiseless scattering states can appear even in the presence of
%diffractive quantum point contacts at the entrance or exit, mediating the
%quantum-to-classical crossover in clean systems. They are, however, already
%strongly suppressed by a weak bulk disorder potential. 
The observation that
$I(T)$ depends on the specific character 
of the diffractive scattering (''bulk vs.~surface disorder'')
is in line with recent
investigations.\cite{falko,sukhorukov,jacquod,aigner,feist}
To the best of our knowledge, the present results explicitly 
demonstrate for the first time in a
genuine 2D system, how these different noise sources influence the
emergence of noiseless scattering states. 

To analyze our findings quantitatively, we compare
our numerical results for the eigenvalue distribution $P(T)$ with the
analytical predictions of Eqs.~(\ref{cross1},\ref{cross2}). Note, however, 
that in our cavities there will always be contributions from ballistic
{\it and} stochastic scattering sources, rather than from either source
alone. {\it Stochastic} scattering events do occur for any variation of
the potential on a scale smaller than $\lambda_F$, such as sharp cavity 
openings or short-range bulk disorder. {\it Ballistic} broadening of 
wave packets is induced by scattering at the cavity walls which, except for
the openings, are always chosen to be suffienctly ``smooth''.
Particularly in the regime of weak disorder both mechanisms will leave
their signatures on the transmission eigenvalue distribution. To properly
account for these signatures we propose to merge the crossover models 
Eqs.~(\ref{cross1},\ref{cross2}) in the following way:
We start from the crossover model for a ballistic
system, $P^{\alpha}(T)$ [Eq.~(\ref{cross1})], which correctly describes 
the appearance of noiseless channels in the absence of bulk disorder 
(Fig.~\ref{fig:1}a). 
Introducing disorder is expected to affect (a) the 
noisy as well as (b) the noiseless part of $P^{\alpha}(T)$ and
will furthermore induce (c) flux exchange processes between these two components.
The effect of the disorder on (a) and (b) is suggested by
Eq.~(\ref{cross2}): Whereas the indeterministic, intrinsically noisy channels 
$P_{RMT}(T)$ in (a) should remain unchanged, the deterministic 
distribution $P_{cl}(T)$ in (b) is expected to 
evolve as described by $P^{\beta}(T)$ in Eq.~(\ref{cross2}). 
The flux exchange between the two phase-space components (c) 
is stochastic and should mutually balance.
This suggests the following crossover model for cavities with 
both ballistic and stochastic scattering sources,
\begin{equation}
P^{\alpha,\beta} (T) = \alpha\,P_{RMT}(T) +
(1-\alpha)\,P^\beta(T)\,.\label{cross3}
\end{equation}
This ''hybrid'' crossover model should serve as a good starting point for analyzing
the case of weak disorder scattering,\footnote[3]{We expect corrections to our 
phenomenological model, Eq.~(\ref{cross3}), to become important for intermediate to 
large disorder strength, where a separation of phase space into two distinguishable
components breaks down.} allowing us to quantify the crossovers
(Fig.~\ref{fig:1}a,b) in the 2D parameter space of  
$\alpha, \beta \in(0,1)$ (see Fig.~\ref{fig:2}). Note that the {\it classical} 
(i.e.~deterministic) limit in this 2D-space corresponds to the point 
$(\alpha\!=\!\beta\!=\!0)$, whereas the {\it quantum} or 
{\it RMT} limit is represented by the lines of the 
parameter space at $\alpha=1$ (ballistic ``quantum chaos'')
and at $\beta=1$ (stochastic ``quantum disorder'').
%(see Fig.~\ref{fig:2})
For the shot noise Fano factor $F$ our hybrid model translates to the 
crossover $F\approx (1/4)\times(1-\alpha\ln\beta)/(1-\ln\beta)$,
thereby reproducing  $F\approx \alpha/4$ 
in the absence of stochastic scattering\cite{agam} ($\beta\rightarrow 0$)
and $F\approx (1/4)/(1-\ln\beta)$ in the absence of ballistic 
scattering\cite{sukhorukov} ($\alpha\rightarrow 0$). Fitting our numerical 
results for $P(T)$ (Fig.~\ref{fig:1}a,b) by Eq.~(\ref{cross3}) 
allows to describe the crossover in terms of different trajectories in the
$(\alpha, \beta)$ parameter space. In the absence of
bulk disorder, the trajectory for decreasing shutter
openings (i.e.~increasing dwell time) features small $\beta$ as it approaches 
the RMT limit. On the other
hand, increasing the disorder potential $V_0$ 
results in the approach of the RMT
limit through a rapid increase in $\beta$ while $\alpha$ tends to zero
(indicating the mergence of the two separate phase space components (a),(b)  
for increasing $V_0$). Note
that with Eq.~(\ref{cross3}) we can directly quantify the signatures 
that either ballistic or stochastic scattering in 2D cavities leave
on $P(T)$. 

At this point the question suggests itself, whether the above differences
in the crossover behavior leave clear signatures in any of the current cumulants
$\langle\langle I^m\rangle\rangle$ that might be accessible experimentally.
For ``symmetric'' cavities with an equal number of incoming and outgoing
channels we 
have found already previously\cite{aigner} that such differences are
hard to pin down in the shot noise (i.e.~in the second cumulant, $m=2$).
A straigtforward evaluation of $\langle\langle I^m\rangle\rangle$ for all $m$ [using
our numerical data from Fig.~\ref{fig:1}a,b or, alternatively,
Eqs.~(\ref{cross1},\ref{cross2})], reveals that the above differences in the crossover
do not lead to characteristic signatures in {\it any} of the {\it individual} current
cumulants. Rather than appearing in individual cumulants explicitly, the characteristic 
differences in the values of $P(T)$ near $T\approx 0,1$ seem to be distributed
over all even cumulants of the current (the odd cumulants are strongly suppressed due to the 
symmetry of $P(T)$ with respect to $\langle T\rangle\approx 0.5$). 
A possible strategy to circumvent this limitation would be to
resort to cavities with different degrees of opening to the
left and right reservoir.\cite{blanter3,sukhorukov,bulashenko2} 
In such ``asymmetric'' cavities 
the internal cavity dynamics is expected to leave clear
signatures already on the third cumulant, a quantity which recently could be
accessed experimentally in tunnel junctions.\cite{reulet}

\begin{figure}[!b]
      \includegraphics[angle=0,width=85mm]{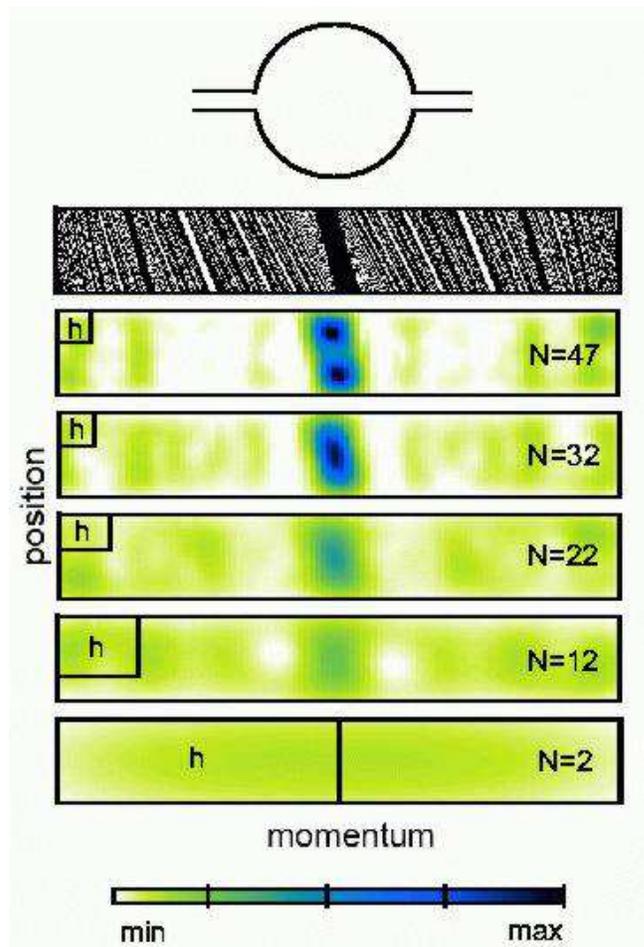}
      \caption{(Color online) Circular billiard (no disorder). Top bar:
        Classical Poincar\'e surface of section
        (transmitted/reflected trajectories represented by
        black/white regions). Bottom bars: 
        Cumulative Husimi distributions $H(x,p)$ 
        of strongly transmitted scattering states (Eq.~\ref{husi}).
        $H(x,p)$ is shown for specific mode intervals 
        $N$ and black frames indicate the size of the Planck cell $h$.
        For $N\gtrsim24$ the largest transmission band (see central black
        region in the Poincar\'e surface) is larger than $h$ and can be resolved
        by the quantum scattering process. Above this threshold, noiseless scattering 
        states appear in $H(x,p)$ in
        form of pronounced density enhancements near the largest 
        transmission bands.}
        \label{fig:3}
\end{figure}

We now probe for the influence of the underlying chaotic classical
dynamics on the transmission eigenvalue statistics at vanishing bulk 
disorder (following previous investigations on shot 
noise\cite{sim,naz,marconcini,aigner}). To this end we contrast the transport
properties of a circular and a stadium billiard (see insets
Figs.~\ref{fig:1}c,d). Due to the classical scaling invariance 
of ballistic billiards with constant potential in the interior,
we can probe the quantum-to-classical crossover, 
$\hbar_{\rm eff}\rightarrow 0$, by the limit 
$k_F\rightarrow\infty$. Although numerically very demanding,
we study the regime \footnote[2] {The cavity area $A=4+\pi$, and $d/\sqrt{A}\approx 0.09$ for both the
circular and the stadium shaped cavity. With this choice of cavity sizes the stadium
with a quadratic rectangular part features two half-circles on either side with radius $r=1$ (in arbitrary length units).}
of comparatively long dwell time $\tau_D$ 
to assure a sufficiently ``universal'' behavior.
For the circular billiard we can reach 50 open lead modes, whereas for 
technical reasons only half as many modes can be accessed for the stadium.\cite{rotter1} 
Both geometries feature a fourfold symmetry, for which case
$P_{RMT}$ [Eq.~(\ref{prmt})] applies also for low mode numbers\cite{gopar} $N$.
Deviations from $P_{RMT}$ can therefore be interpreted as contributions of
noiseless channels. 

% Any deviations from 
%$P_{RMT}$ should therefore originate
%from the system-specific dynamics in the regular cavity or from contributions
%of noiseless channels.
%In each mode interval, we calculate the scattering matrix at $200$
%aequidistant points in k-space. In the circular structure, we reach an energy
%corresponding to $50$ open modes in the lead, in the stadium-shaped structure, which is
%numerically more challenging, we reach an energy corresponding to $25$ open modes.
 
Remarkably, for the low-energy interval $k_F\!\in\![1, 5] \pi /d$, 
we find excellent agreement between the numerical
distribution $I(T)$ and its RMT-prediction, for
{\it both the stadium and the circle billiard} (see Fig.~\ref{fig:1}c,d).
Differences between regular and chaotic 
dynamics do not leave any imprint on $I(T)$. At higher electron energies
(or smaller wavelengths) the onset of noiseless 
scattering is similarly reflected in $I(T)$ for both
geometries (Fig.~\ref{fig:1}c,d),
irrespective of the classical chaoticity or the lack thereof. 
This finding points to the conclusion that 
the appearance of the first noiseless states is uniquely determined by the
requirement that quantum mechanics can
resolve the largest classical transmission band in phase space.
\cite{silvestrov,jacquod} Since both geometries
feature the same lead width $d$, which controls the size of these %first direct 
transmission bands, noiseless states
should appear at approximately the same $k_F$ for both 
cavities.

\begin{figure}[!t]
      \includegraphics[angle=0,width=85mm]{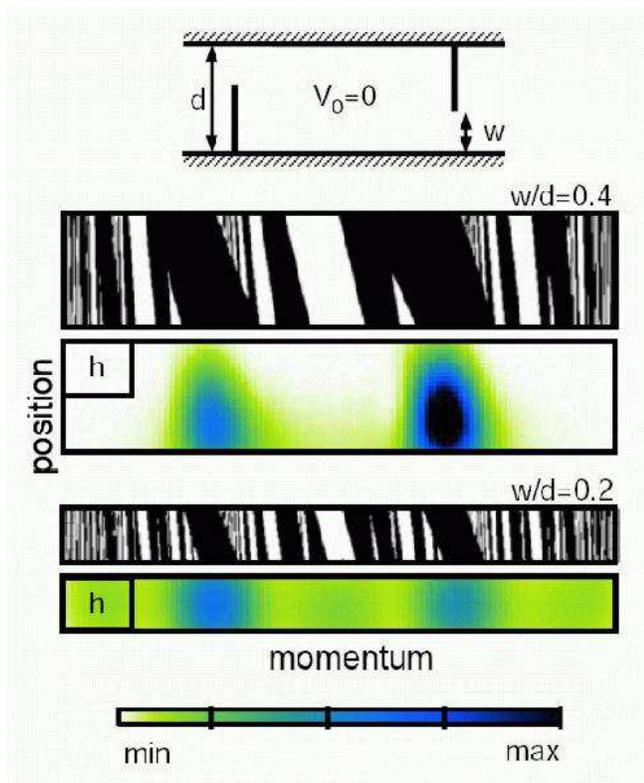}
      \caption{(Color online) Rectangular billiard with tunable opening
        (no disorder, same color coding as in Fig.~\ref{fig:3}). Top two bars: 
        Large shutter openings, $w/d = 0.4$. Bottom two bars: 
        Weakly open shutters, $w/d = 0.2$.
        Decreasing the shutter opening below the threshold value 
        $w/d\approx 0.32$ (where
        the size of the Planck cell $h$ is equal to the largest transmission 
        band) reduces
        any pronounced enhancements in the Husimi distributions $H(x,p)$.}
        \label{fig:4}
\end{figure}

To further investigate this issue we demonstrate the quantum resolution 
of the classical phase space explicitly. To this end we compare
Husimi distributions of scattering states with the Poincar\'e surface of 
section recorded at the entrance lead mouth.\cite{tworz,wein} 
We calculate the cumulative Husimi function containing those eigenstates 
$|T_i\rangle$ of $t^\dagger t$ which correspond to the 
largest transmission eigenvalues $T_i$ within a given energy interval,
\begin{equation}\label{husi}
H(x,p)=\sum_i^{M}H_i(x,p)=\sum_i^{M}
|\langle T_i|x, p \rangle|^2\,.
\end{equation}
$|x, p \rangle$ is a coherent state of minimum uncertainty with its peak
at the position $x,p$ and the number of eigenstates $|T_i\rangle$ that 
contribute to the above sum is chosen as $M=2N$.
In line with our calculations for the integrated eigenvalue distribution 
$I(T)$ (see Fig.~\ref{fig:1}), we now probe how electron energy, cavity opening, 
and disorder strength affect the distribution $H(x,p)$.
In the circular billiard we probe the quantum-to-classical 
crossover by evaluating $H(x,p)$ in specific mode intervals $N$, corresponding to
different electron energies (200 equidistant energy 
points per mode interval are calculated).
We find that for low mode numbers $N$, the distribution $H(x,p)$ covers large 
parts of phase space more or less uniformly (see Fig.~\ref{fig:3}). For higher mode numbers,
$H(x,p)$ shows a drastic enhancement 
near the largest transmission bands in phase space and a strongly reduced amplitude 
elsewhere. Comparing the size (area) of the Planck cell $h$ 
(indicated by the black frames in Fig.~\ref{fig:3}) to that of the largest 
transmission band (see the central black region in the classical Poincar\'e surface), 
we obtain an estimate for
the threshold value above which noiseless scattering should appear. For the
circular billiard this threshold is given by $k_F\approx 24\pi/d$ (i.e.~$N\approx 24$), at which
value the largest transmission band and $h$ become equal in 
size.\footnote[1]{Note that in the stadium billiard this threshold value is somewhat
larger than in the circle ($k_F\approx 32\pi/d$), due to the larger geometrical distance
between lead mouths in this geometry.} This estimate indeed accurately predicts, above
which value of $N$ our numerical results for $H(x,p)$ (Fig.~\ref{fig:3}) show significant
enhancements near the largest
transmission bands. 

\begin{figure}[!t]
      \includegraphics[angle=0,width=85mm]{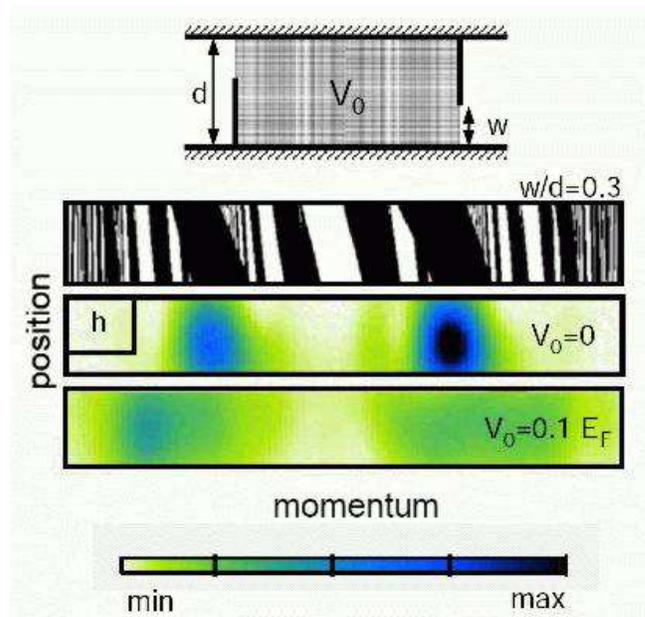}
      \caption{(Color online) Rectangular billiard with tunable bulk disorder
        (fixed opening ratio $w/d=0.3$, same color coding as in
        Fig.~\ref{fig:3}).
        Top two bars: No disorder. Bottom bar: Moderate disorder, $V_0=0.1
        E_F$.
        Bulk disorder destroys the appearance of noiseless states due to
        stochastic scattering.}
        \label{fig:5}
\end{figure}

We now perform a similar analysis for the 
tunable rectangle. For this cavity we keep the 
electron energy fixed in the averaging interval $k_F\in [40.1, 40.85] \times \pi /d$
and vary the cavity opening $w$ or, alternatively, the disorder strength $V_0$.
By tuning the cavity openings at fixed energy we change the 
size of the largest transmission band at a fixed value of $h$. When these 
two phase space areas are equal in size, we obtain a threshold value  
for the appearance of the first noiseless state in terms of the cavity opening:
$w/d\approx 0.32$. Comparing this estimate with our numerical results for $H(x,p)$
(see Fig.~\ref{fig:4})
yields again very good agreement: Whereas for an opening of $w/d=0.2<0.32$ the Husimi
function $H(x,p)$ looks rather flat (Fig.~\ref{fig:4} bottom), very clear enhancements around the 
largest transmission band appear for $w/d=0.4>0.32$ (Fig.~\ref{fig:4} top).
In Fig.~\ref{fig:5} we demonstrate that bulk disorder in the
cavity destroys any noiseless states by strongly reducing any
pronounced enhancements which would otherwise be present in $H(x,p)$. 
We finally note that in all of the above cases (Figs.~\ref{fig:3}-\ref{fig:5}) 
drastic enhancements in $H(x,p)$ always come along with a corresponding offset in the
integrated eigenvalue distribution $I(T)$ and vice versa. This evidence should unambiguously
document the presence of noiseless scattering
states in 2D-cavities. 

To summarize, we have identified signatures of ballistic and
stochastic scattering in the quantum-to-classical crossover 
of clean and disordered samples. 
A model for the transmission eigenvalue distribution $P(T)$ is proposed which 
combines previous approaches for the ballistic and disordered 
limit\cite{jacquod,sukhorukov} and which allows to extract contributions of different
noise sources to our numerical results for $P(T)$.
We provide the first evidence for ``noiseless scattering
states'' \cite{silvestrov} in clean, genuine 2D-cavities and confirm the corresponding
decomposition of the electronic flow in a classical and a quantum component.
\cite{jacquod} The emergence of noiseless states is found to be 
determined by the size of the largest classical transmission band\cite{wirtz1} 
in phase space. The latter quantity,
in turn, depends on the system specific geometry of the cavity and not 
necessarily on its 
chaoticity or on the lack thereof. In the presence of bulk disorder, noiseless scattering
states disappear due to stochastic scattering, as previously anticipated.\cite{sukhorukov} 

We thank E.V.~Sukhorukov for useful suggestions 
and V.~A.~Gopar, H.~Schomerus, A.~D.~Stone, B.~Weingartner
for helpful comments and discussions.  Support by the FWF-Austria
(Grants Nos.~FWF-SFB016 and FWF-P17359) and by 
the Max-Kade Foundation (New York) are gratefully acknowledged.

\end{document}